\documentclass[referee]{raa}            % referee version: for submission

\usepackage{graphicx,times,epsf} %for PS/EPS graphics inclusion, new
\usepackage{geometry} 
\begin{document}

   \title{Search for Giant Pulses of radio pulsars at frequency 111 MHz with LPA radio telescope
%\,$^*$
%\footnotetext{$*$ Supported by the National Natural Science Foundation of China.}
}
%   \subtitle{I. Place Your Subtitle Here}

   \volnopage{Vol.0 (200x) No.0, 000--000}      %%preserved for Editor. DOn't remove!
   \setcounter{page}{1}          %%starting page, preserved for Editor. DOn't remove!

   \author{A. N. Kazantsev
      %\inst{1}
   \and V. A. Potapov
      %\inst{1}
   %\and B. J. Smith
    %  \inst{3}
   }
%% Here is an example of three authors come from different institutes.
%% For single author or all the authors from an institute, use "\inst{}" only

   \institute{P.N. Lebedev Physical Institute of the Russian Academy of Sciences, Pushchino Radio Astronomy Observatory,
             Pushchino 142290, Russia; {\it kaz.prao@bk.ru}\\
%% Please give the E-mail address of the author, to whom future correspondence and
%% offprint requests will be sent.
        %\and
         %    Full institute address for the second author\\
        %\and
         %    Full institute address for the third author\\
   }

   \date{Received~~2017 month day; accepted~~2017~~month day}

\abstract{ We have used the unique low frequency sensitivity of the Large Phased Array radio telescope of Pushchino Radio Astronomy Observatory to collect dataset of the single pulse observations of second period pulsars of the Northern hemisphere. During observational sessions of 2011 - 2017 yrs. We have collected data on 71 pulsars at 111 MHz frequency using digital pulsar receiver. We have discovered Giant Radio Pulses (GRP) from pulsars B0301+09 and B1237+25, and confirmed early reported generation of anomalously strong (probable giant) pulses from B1133+16 on statistically significant dataset. Data for these pulsars and from B0950+08, B1112+50, early reported as pulsars generating GRPs were analyzed to evaluate their behavior on long time intervals. It was found that statistical criterion (power-law spectrum of GRPs distribution on energy and peak flux density) seems not to be strict for pulsars with the low magnetic field on light cylinder. Moreover, spectra of some of these pulsars demonstrate unstable behavior with the time and have complex multicomponent shape. In the dataset of B0950+08 we have detected strongest GRP from pulsars with the low magnetic field on light cylinder ever reported having the peak flux density as strong as 16.8 kJy.
}

   \authorrunning{A. N. Kazantsev \& V. A. Potapov }            %author_head in even pages
   \titlerunning{Search for Giant Pulses of radio pulsars at frequency 111 MHz}  % title_head in odd pages

   \maketitle
%% The author head (on even pages) and the title head (on odd pages) will be
%% automatically extracted from \author{} and \title{}. Whenever the title is too long,
%% you will be asked to supply a shorter one by inserting either \authorrunning{} or
%% \titlerunning{} before \maketitle. Anyway, you can specify your own heads.
%%
%%
%% Note: In the following text body of your manuscript, please note several differences from
%%       other major journals:
%% (1) \subsection{Please Capitalize the First Letter of Each Notional Word in Subsection Title}
%% (2) Please Capitalize the First Letter of Each Notional Word in all tables' captions

%
%________________________________________________ sections below
%
\section{Introduction}  %% first-level sections will be auto-capitalized
\label{sect:intro}

Typical radio pulsars have very stable average pulse profiles, obtained with the summation of the sequence of thousands of individual pulses. At the same time individual pulses emitted by neutron star are very unstable both due instability of the process of pulses emission and effects of the perturbation of the interstellar medium. Individual pulses may be observed for only about one third of known radio pulsars because of their low flux density. Individual pulses can differ essentially from each other by shape, duration and intensity (energy and peak flux density). When shape and duration of individual pulses may change in quite wide limits, fluctuation of intensity can reach extraordinary values. A strong fluctuation of intensity was noticed in an early work about pulsar in Crab nebula (\cite{staelin68}). An accurate analysis of Crab pulsar at the 160 MHz frequency was carried out in \cite{sutton71}. It was shown that Crab pulsar regularly generates very strong individual pulses, and intensity distribution of pulses demonstrates bimodal form: log-normal distribution for quite weak (regular) pulses and power-law distribution for strong pulses. Strong pulses which exceed the average pulse more than 30 times by intensity were called Giant Radio Pulses (GRP). Statistical criterion (power-low form of intensity distribution) became one of the most important for classification of GP. For a long time Crab pulsar was the only known pulsar emitting GRPs. Later on there were found GRPs from the fastest known (at the time) millisecond pulsar B1937+21 \cite{wolszczan84}, \cite{cognard96}. After many years there were discovered similar phenomena from set of pulsars (\cite{singal01}; \cite{romani01}; \cite{ershov03}; \cite{joshi04}; \cite{kuzmin04}; \cite{joshi04}; \cite{knight05}; \cite{kuzmin06}; \cite{ershov06}). Discovery of GRPs from pulsars out of our Galaxy was reported for two pulsars in the large Magellanic cloud (\cite{johnston03}; \cite{crawford13}). 

List of criteria of giant pulses which were formed over time contains the following items:

\begin{itemize}
\item GRP peak flux density is 30 times or more than strong and the energy in individual pulse is more than 10 times as strong as these of an average pulse;
\item GRPs are localized at longitude of average pulse profile or interpulse;
\item peak flux density is about hundreds and thousands (in cases of Crab pulsar even millions) Jy;
\item short duration of GRP in comparison with an average pulse and an extremely short duration of it's microstructure components (up to 0.4 ns for Crab pulsar \cite{hankins07});
\item high level of linear and circular polarization of GRP;
\item power-law peak flux density and energy in pulse distribution;
\item extremely high brightness temperature for narrower pulses, up to about $10^{41}$ K for Crab pulsar \cite{hankins07})
\end{itemize}

Not all itemized above parameters presents at the same time for all GRPs. Sometimes it that makes a bit tricky the strict definition of the GRP. Mostly statistical criterion (connected with the type of distribution) is taken as crucial, but, as it will be shown below, this criterion seems to be not so robust as it was as was thought previously, at least for seconds period pulsars with the low magnetic field on light cylinder.

It worth to note that generation of the strong pulses more than 30 times of an average is very rear for the vast majority of pulsars. In case, we can see such a phenomenon it's mostly connected with the far right "tail" of log-normal distribution in flux (\cite{taylor71}; \cite{hesse74}; \cite{ritchings76}).

It is easy to see that the pulsars in table can be divided into two subclasses:

\begin{itemize}
\item Pulsars with strong (over $B_{LC} > 10^{5}$ G) magnetic fields at their light cylinders and millisecond period of their rotation that can be observed
at frequencies above 600 MHz. The most typical pulsars of this subclass are B0531+21 Crab Nebula pulsar and the rapid millisecond pulsar B1937+21 were observed over wide ranges of frequencies. Their pulses demonstrate submicrosecond sub-structure and extremely high peak flux density up to millions Jy.
\item Pulsars with $B_{LC}$ from several to several hundred Gauss. These are pulsars with second periods that have been observed only at frequencies 40 -- 111 MHz (with the only exception of J0529-6652). Their GRPs are as strong as several hundreds or thousands Jy, and exceed the average profiles by factors of tens and their widths are of the same order as the average profile.
\end{itemize}

The problem of an adequate theoretical description of the phenomenon of GRP generation faces with a wide range of parameters of pulsar with GRPs. List of such pulsar includes both milliseconds pulsars and second period pulsars, pulsars with weak and large magnetic field on light cylinder, young active pulsar (as Crab pulsar) and old recycled pulsar (as B1937+21). Up to now no one theoretical model have been made which can predict emission of GRPs by using function of known pulsar parameters. The majority of existing models related to description of, so called, “classical” pulsars with GRPs (B0531+21 Crab pulsar and B1937+21) (\cite{petrova06}). Existing deficit of the observed data for second period pulsars with low magnetic field on light cylinder in comparison with these for millisecond pulsars encouraged us to implement a long-term observational program for pulsars of Northern hemisphere both to search for the new pulsars with GRPs and monitoring of known pulsars with GRPs.

Here we present some recent results of this search and monitoring program of GRPs that have been making with Large Phase Array in 2011-2017.

\begin{table}
\begin{center}
\caption{List of Pulsars with GRPs. }
\bigskip
\begin{tabular}{lcccl}
\hline
Pulsar name           &  Period     & Frequency  &  $B_{LC}$                     & First   \\
(epoch 2000 and 1950) &  [s]        & [MHz]      &   [G]                         & published \\
\hline
J0034-0721 (B0031-07) &   0.9429      &      40    &   7.02                      &  \cite{kuzmin04}   \\
J0218+4232            &   0.0023      &     610    &   $31.21 \times 10^{5}$     &   \cite{joshi04}    \\
J0304+1932 (B0301+19) &   1.3876      &     111    &   4.76                      &   \cite{kazantsev17a} \\   
J0534+2200 (B0531+21) &   0.0331      &  40-8300   &   $9.8 \times 10^{5}$       &   \cite{staelin68} \\
J0529-6652*           &   1.0249      &     610    &   39.7                      &  \cite{crawford13} \\
J0540-6919*           &   0.0505      &    1390    &   $3.62 \times 10^{5}$      &   \cite{johnston03} \\
J0659+1414 (B0656+14) &   0.3849      &    111     &   766                       &  \cite{kuzmin06} \\
J0953+0755 (B0950+08) &   0.2530      &    111     &   141                       &  \cite{singal01} \\
J1115+5030 (B1112+50) &   1.6564      &    111     &   4.24                      &  \cite{ershov03} \\
J1136+1551 (B1133+16) &   1.1879      &    111     &   1.19                      &  \cite{kazantsev15b} \\
J1239+2453 (B1237+25) &   1.3824      &    111     &   4.14                      &  \cite{kazantsev15a}  \\
J1752+2359            &   0.4091      &    111     &   71.1                      &  \cite{ershov06} \\
J1824-2452A           &   0.0030      &   1510     &   $7.41 \times 10^{5}$      &   \cite{johnston03}  \\
J1823-3021A           &   0.0054      &    685     &   $2.52 \times 10^{5}$      &   \cite{knight05}  \\
J1939+2134 (B1937+21) &   0.0016      &  111-5500  &   $1.02 \times 10^{6}$      &   \cite{wolszczan84} \\
J1959+2048            &   0.0016      &     610    &   $3.76 \times 10^{5}$      &   \cite{joshi04} \\
\hline
\multicolumn{5}{l}{Period of pulsar is shown up to 4 digits after the decimal.} \\
\multicolumn{5}{l}{$B_{LC}$ is the magnetic field on the light cylinder in Gauss} \\
\multicolumn{5}{l}{* pulsars in LMC.}
\end{tabular}
\end{center}
\label{intro-psrdata}
\end{table}

\section{Observations and data reduction}
\label{sect:observ}
The observations were carried out in 2011 -- 2017 using the 1st (one beam) diagram of Large Phased Array (LPA) transit radio telescope of Pushchino Radio Astronomy Observatory (Astro Space Center, Lebedev Physical Institute). The telescope has an effective area in the zenith direction of about 20000$\pm$1300 $m^{2}$. One linear polarization was used. A 512-channel digital receiver with synthesized channels with bandwidth $\Delta$f = 5 kHz each was used. The full bandwidth of observations was 2.3 MHz (460 frequency channels were used) and was reduced by amplitude-frequency response of the analog receiver. The main frequency of the observations was 111 MHz. The sampling interval was 1.2288 ms for the majority of observations, and several sessions of observations were carried out with 2.4576 ms sampling. Duration of observation session was taken equal to the time of passage of the pulsar through 0.5 of antenna diagram and did depend on the declination of pulsar, typical it was between 3 and 10 minutes (about 3.5 min for PSR B1237+25 or is about 11 min for PSR B0809+74, for example). All observations were carried out in individual pulses mode of digital receiver (sequence of individual pulses of pulsar were recorded).

Phased analog signal from radio telescope's dipoles is served to the receiver. A synchronizer generates trigger pulses that synchronize the work of the digital receiver with a period of given pulsar. Precise timing is provided by signals from the GPS. The accuracy of the converting GPS time to the universal time scale is $\pm$~100 ns. The accuracy of setting the start time of the receiver by the synchronizer is $\pm$~10 ns, that far exceeds any reasonable requirements for accuracy of observations at 111 MHz frequency. For each trigger pulse, the signal is digitized at a frequency of 5 MHz and accumulate in the receiver's buffer. These readings from the buffer are loaded into the fast Fourier transform hardware processor. Digitization of the input signal and filling of the buffer are conducted continuously without interruptions in time. Thus, continuous generation of input signal spectra is made, each of which contains 512 spectral channels.

\begin{figure}
   \centering
   \includegraphics[width=9cm,angle=0]{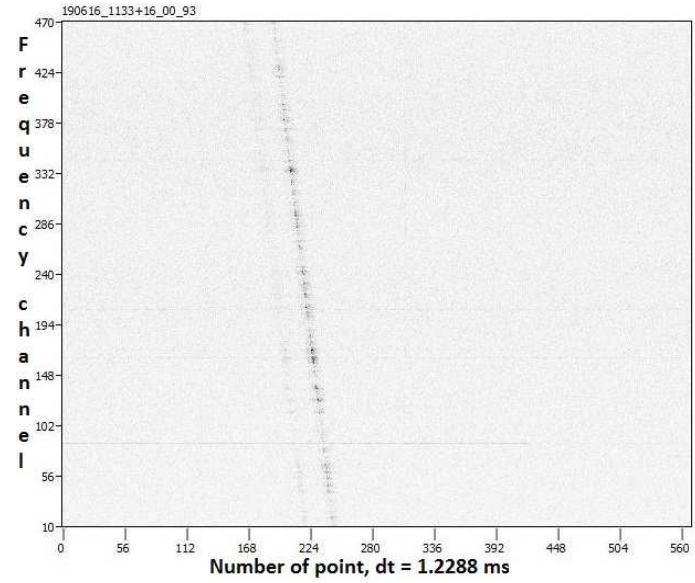}
   \caption{Example of dynamical spectrum for individual pulse of B1133+16 in digitally synthesized 512 frequency channels with uncompensated dispersion delay.}
   \label{fig1}
\end{figure}

A resulting file of an observation in the individual pulses mode consists of a header and array of time-series spectra. Data of time-series spectra are recorded as 32-bit floating-point numbers. So, the data looks like 3D array: frequency (in frame of receiver’s bandwidth), point(time), intensity (in analog-digital converter units). Plot of individual pulse data in frequency-time coordinates clearly demonstrates dispersion delay of pulsar’s pulse (see Fig.~\ref{fig1}).  For every pulse the dispersion delay was compensated off-line. As a rule, such a procedure effectively eliminates impulse interference from Earth sources. In this way, sequences of compensated by dispersion measure (DM) to the 111 MHz frequency individual pulses was collected as primary information for a subsequent analysis.

\begin{table}
\begin{center}
\caption[]{Information About Observations for Each Pulsar. DM Denotes Disperion Measure of Interstellar Medium in the Direction to the Given Pulsar.}\label{Tab:publ-works}

%%Please Capitalize the First Letter of Each Notional Word in table's caption

\begin{tabular}{clclcl}
  \hline\noalign{\smallskip}
No &  Name of pulsar      & Period     & DM               &  Number of observ. sessions &  Total time of data folding  \\
   &  epoch 1950(2000)    & [s]        & [$pc \ cm^{-3}$] &                             &  [hour]  \\
  \hline\noalign{\smallskip}
1  & B0301+19 (J0304+1932) & 1.3876     & 15.66  & 275  & 15.47\\ 
2  & B0320+39 (J0323+3944) & 3.0321     & 26.19  & 155  & 10.70 \\
3  & B0329+54 (J0332+5434) & 0.7145     & 26.76  & 63   & 5.80  \\
4  & B0809+74 (J0814+7429) & 1.2922     & 5.75   & 588  & 116.72 \\
5  & B0950+08 (J0953+0755) & 0.2531     & 2.97   & 677  & 36.40   \\
6  & B1112+50 (J1115+5030) & 1.6564     & 9.19   & 1063 & 89.01    \\
7  & B1133+16 (J1136+1551) & 1.1879     & 4.84   & 616  & 33.94     \\
8  & B1237+25 (J1239+2453) & 1.3824     & 9.25   & 568  & 33.37       \\
  \noalign{\smallskip}\hline
\end{tabular}
\end{center}
\end{table}

\begin{figure}[h]
  \begin{minipage}[t]{0.495\linewidth}
  \centering
   \includegraphics[width=75mm,height=61mm]{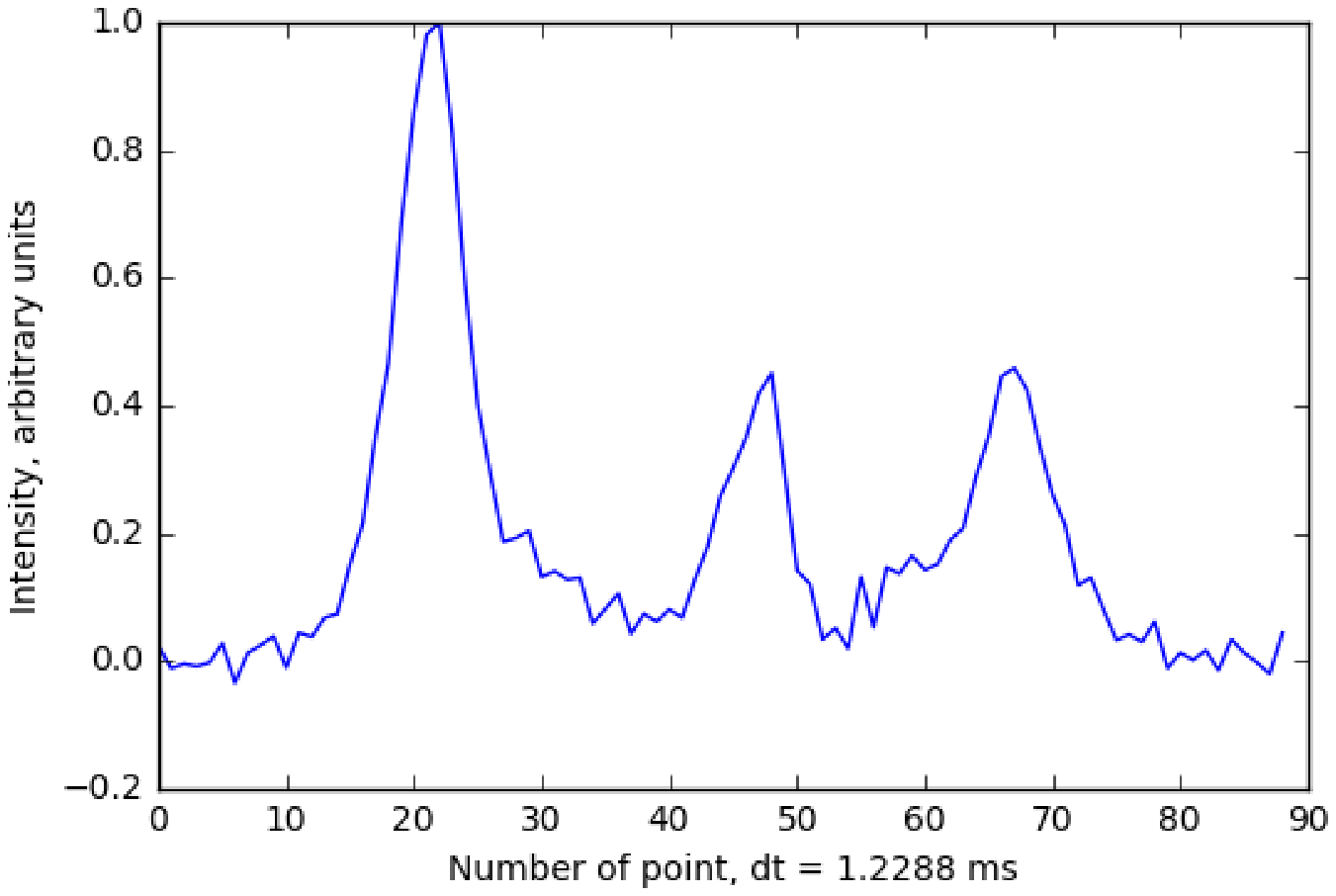}
   \caption{{\small Pattern average pulse profile of B1237+25.} }
  \label{fig2}
  \end{minipage}%
  \begin{minipage}[t]{0.495\textwidth}
  \centering
   \includegraphics[width=75mm,height=61mm]{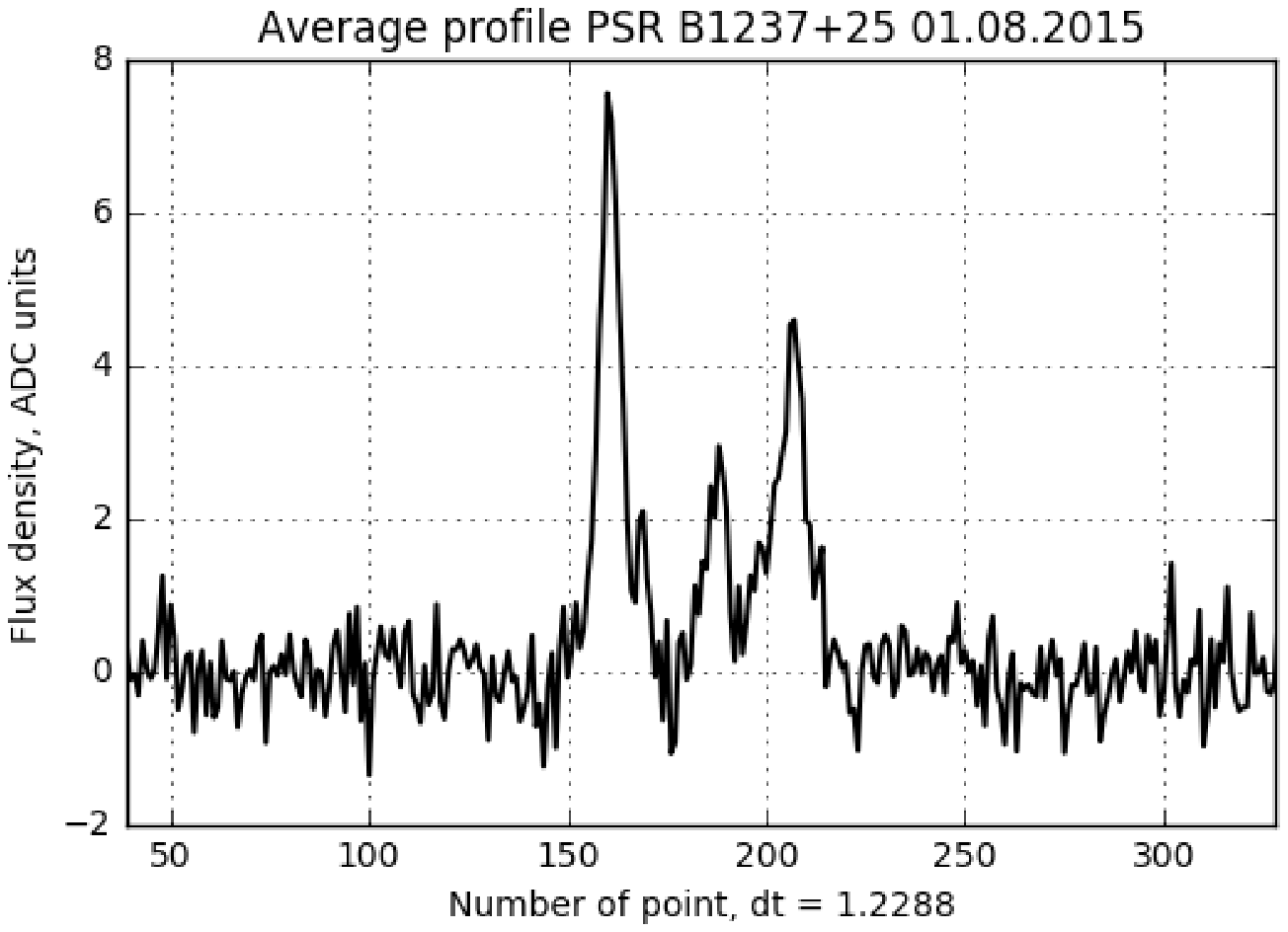}
  \caption{{\small Dynamical average pulse profile of B1237+25.}}
  \label{fig3}
  \end{minipage}%
\end{figure}

\section{Data analysis}
\label{sect:analysis}

The data of sequences of individual pulses were used to generate a dynamical average pulse profile (average pulse profile per one observational session).  Longitude of an average pulse profile and off-pulse region were determined by correlation with a pattern average pulse profile of every pulsar.  (see Fig.~\ref{fig2} and Fig.~\ref{fig3}). Off-pulse region was used to calculate background noise $\sigma_{noise}$. Then, pulses with a phase on a longitude of an average pulse profile of pulsar and a peak flux density more than 4$\sigma_{noise}$, were collected. For each pulse, intensity in ADC units, intensity in peak flux density of a dynamical average pulse profile, and phase in sample points were calculated. All data which have satisfied foregoing conditions were combined in one data array. The resulting data were adjusted for the telescope’s beam shape both for azimuth and zenith distance. Then search for strongest pulses with peak flux density more than 30$S_{avg}$ (peak flux density of a dynamical average pulse profile of a pulsar) was made. Data were calibrated using a noise signal (noise step) with a temperature of 2100 K injected into the receiver tract before amplifier that enables to estimate the flux density in Jy as $S = 2kT /A_{eff}$, where $S$ is the peak flux density, $T$ - the source brightness temperature, and $A_{eff} = 20 000 \pm 1300 m^2$ is the effective area of the 1st diagram of LPA.  Such a procedure gives an estimate of the total rms uncertainty in absolute flux density units in about 18 \% (\cite{kazantsev17b}). For pulsars B0329+54 and B1133+16 the averaged value for system temperature during observation (5 K for ADC unit) was taken due to the shortage of calibration data, therefore further evaluation of their peak flux density in Jy should be mentioned as an upper limit. Finally, the statistical analysis of pulse’s distributions in absolute (in Jy) and in $S_{avg}$ per session units was made for pulsars with GRPs to prove statistical (power-low) criterion.

\section{Results}
\label{sect:results}
\subsection{Pulsars without GRPs}

As it is well-known, that the overwhelming majority of pulsars do not generate GRPs. Nevertheless, it was interesting for us to obtain a pattern of statistic for set of such pulsars to compare it with this of pulsars with GRPs. This sub-section includes data on three of observed pulsars which didn't demonstrate GRPs or giant-like pulses. These pulsars are referred further as "regular" or "normal" radio pulsars.

\subsubsection{B0320+39 (J0332+5434)}

B0320+39 is slow rotating pulsar of the Northern hemisphere. The pulsar has two-componets average profile at the 111 MHz. Notices about giant pulses from this pulsar were not ever reported. There were detected and analyzed 9406 pulses which satisfied the conditions from \ref{sect:analysis}. The strongest observed pulse is 29 times larger than the dynamical average pulse profile (see Fig.~\ref{fig_pulse}~(a)) and has width 1/4 times as narrow as an average pulse at 10\% of average pulse peak flux density and about 1/2 of relevant component at 50\%. Figure~\ref{fig_disp}~(a) shows that strong pulse is well prominent among regular pulses.  Pulse intensity histogram (see Fig.~\ref{fig_hist}~(a)) and distributions have the form close to the simple log-normal (power-law) distribution.

\subsubsection{B0329+54 (J0332+5434)}

B0329+54 is one of the brightest isolated pulsar which, possible, have a planetary system (\cite{shabanova95}), see (\cite{starovoit17}) for discussion. As in the case of B0320+39, there were not detected any GRPs or anomalous pulses for this pulsar. We analyzed 22874 individual pulses which satisfied the conditions. There were no found pulses with peak flux density exceeded this the dynamical average profile more than 30 times. One of the strongest observed pulse is shown in Figure~\ref{fig_pulse}~(b). As we can be seen from the figure, the power pulse has comparable duration with a main component of the average profile of the pulsar and shifted to it's leading edge. In spite of the absence of GRPs, distributions of the flux density both in absolute and relative to dynamical average profile units is quite complicated and can't be described with the simple log-normal model, Fig. \ref{fig_dis_pulse}, \ref{fig_dis_energy}.  
 
\subsubsection{B0809+74 (J0814+7429)}

Because of the high declination of pulsars, duration of one session of observation with LPA was longest and equal about 11 minutes. This allowed us to detect 281331 individual pulses of pulsar which satisfied our conditions. So many pulses make this research the most voluminous for this pulsar. 49 pulses with peak flux density exceeded the peak flux density of the dynamical average profile by more than 30$S_{avg}$ were observed. Figure~\ref{fig_pulse}~(c) shows an example of one strong pulse. For the first time, similar pulses were observed at 18 --- 40 MHz (\cite{ulyanov06}) and were named as anomaly strong pulses. The pulsar has a distinct drifting sub-pulse phenomenon (\cite{page73}) which lead to a tangible "smudging" of peak flux density of average profile. Distributions both in peak flux dencity and in units of $S_{avg}$ for pulses with flux $> 15 S_{avg}$ show combined distribution with power-law part that are not strong enough to classify pulses as GRP.

\subsection{"Old" pulsars with GRPs}
The block includes pulsars with phenomenon of GRPs generation which was discovered earlier outside the frame of our observational program. For these objects a comparative analysis of statistical distribution was done. It should be noted that the discoveries and observations were carried out at the 103-180~MHz that facilitates comparative analysis of the statistics. 

\subsubsection{B0950+08 (J0953+0755)}

For the first time, GRPs from the pulsar were discovered by \cite{singal01} at the 103 MHz. The 20$S_{avg}$ was taken in \cite{singal01} as a threshold for pulses to be called giant. Under this approach, roughly one percent from about one million pulses from B0950+08 were classified as GRPs. We analyzed 155834 individual pulses from pulsar and detected 3422 pulses more than 20$S_{avg}$ (and 1062 pulses more than 30$S_{avg}$), representing 2.2(0.68) percent of pulses with more than 4$\sigma_{noise}$ by peak flux. Observations of giant pulses from the pulsar were carried out using LPA by \cite{smirnova12} earlier. The strongest individual pulses in papers cited above was 300 and 508 times as strong as an average profile, respectively. In frame of present work the strongest pulse in 139.2$S_{avg}$ was detected and is shown in Figure~\ref{fig_pulse}~(d). Notable third-party peaks are interference. At the same time the strongest pulse detected during our observational program has peak flu density 16.8 kJy that is the strongest value ever detected for this pulsar (pulse as strong as 15.2 kJy was detected earlier in (\cite{smirnova12})). It worth to note that distribution of GRPs in our work (Fig. \ref{fig_dis_energy}~(d)) differs significantly from the two-component distribution found in (\cite{smirnova12}) that can be explained instability of distribution on relative short duration of time. 

\subsubsection{B1112+50 (J1115+5030)}
\cite{ershov03} reported about GRPs from this pulsar at a frequency of 111 MHz. In 105 observing sessions 126 GRPs were detected in cited work. This represents about 0.67 per cent of the full quantity of pulsar's periods. In our research this ratio is equal 0.83 per cent. The duration and localization of GRPs are consistent in both works. Sometimes observed GRPs have a complex form which can be formed by a microstructure that can't be resolved at low radio frequencies. Figure~\ref{fig_pulse}~(e) shows one of the giant pulses in comparison with the increased dynamical average pulse profile. Distribution in absolute peak flux density demonstrates power-law behaviour with power index $\alpha = -2.86$ that differs from earlier result of \cite{ershov03} with $\alpha = -3.6$.

\subsubsection{B1133+16 (J1136+1551)}
Bright pulses from the pulsar were noted earlier in (\cite{kramer03}) at 5 GHz but was mentioned as possible giant pulses by \cite{karuppusamy08} and \cite{kazantsev15b}. We analyzed 58642 individual pulses of pulsar B1133+16 and detected 153 pulses more than 30$S_{avg}$. Pulsar has an average profile with two well separate components. GRPs were detected in both. Example of observed giant pulses and the increased dynamical average profile are shown in Figure~\ref{fig_pulse}~(f). The pulse has extremely short duration in comparison of the relevant component average profile of pulsar (unresolved at 1.2288 ms sampling on 50\% of intensity). Distribution of pulses is quite complicated and may be presented for strong pulses as two-component power-low distribution \ref{fig_dis_energy} ~(f) with the bend near 250 Jy.

\subsection{"Fresh" pulsars with GRPs}
The section contains new pulsars with GRPs which were discovered in the frame of our search program with Large Phase Array at the 111 MHz.

\subsubsection{B0301+19 (J0304+1932)}
For the first time, existence of GRPs from the pulsar was reported by \cite{kazantsev17a}. Our previous research included a small amount of sessions of observations. Only 884 pulses with peak flux density more than 3$\sigma_{noise}$ were analyzed. In the present work the sample of 3160 individual pulses of pulsar which were more 4$\sigma_{noise}$ by amplitude was obtained and processed. 80 pulses which can be classified as giant were detected. B0301+19 is enough weak pulsar and it's dynamical average profile is faint except of time of generation of the GRPs, when average profile becomes better detectable.

\subsubsection{B1237+25 (J1239+2453)}
\begin{figure}
   \centering
   \includegraphics[width=15 cm,angle=0]{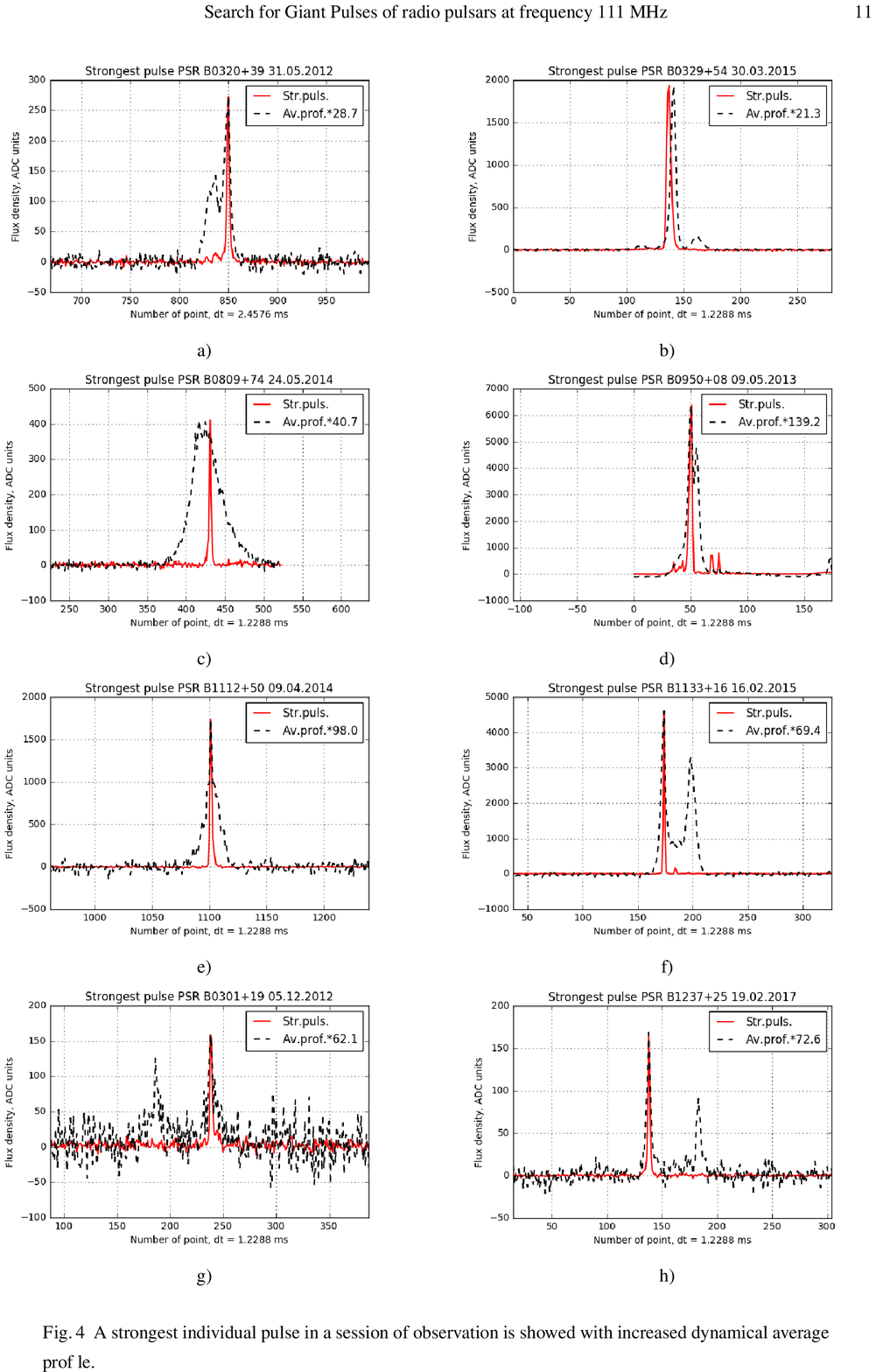}
   \caption{A strongest individual pulse in a session of observation is showed with increased dynamical average profile.}
   \label{fig_pulse}
\end{figure}

\begin{figure}
   \centering
   \includegraphics[width=15 cm,angle=0]{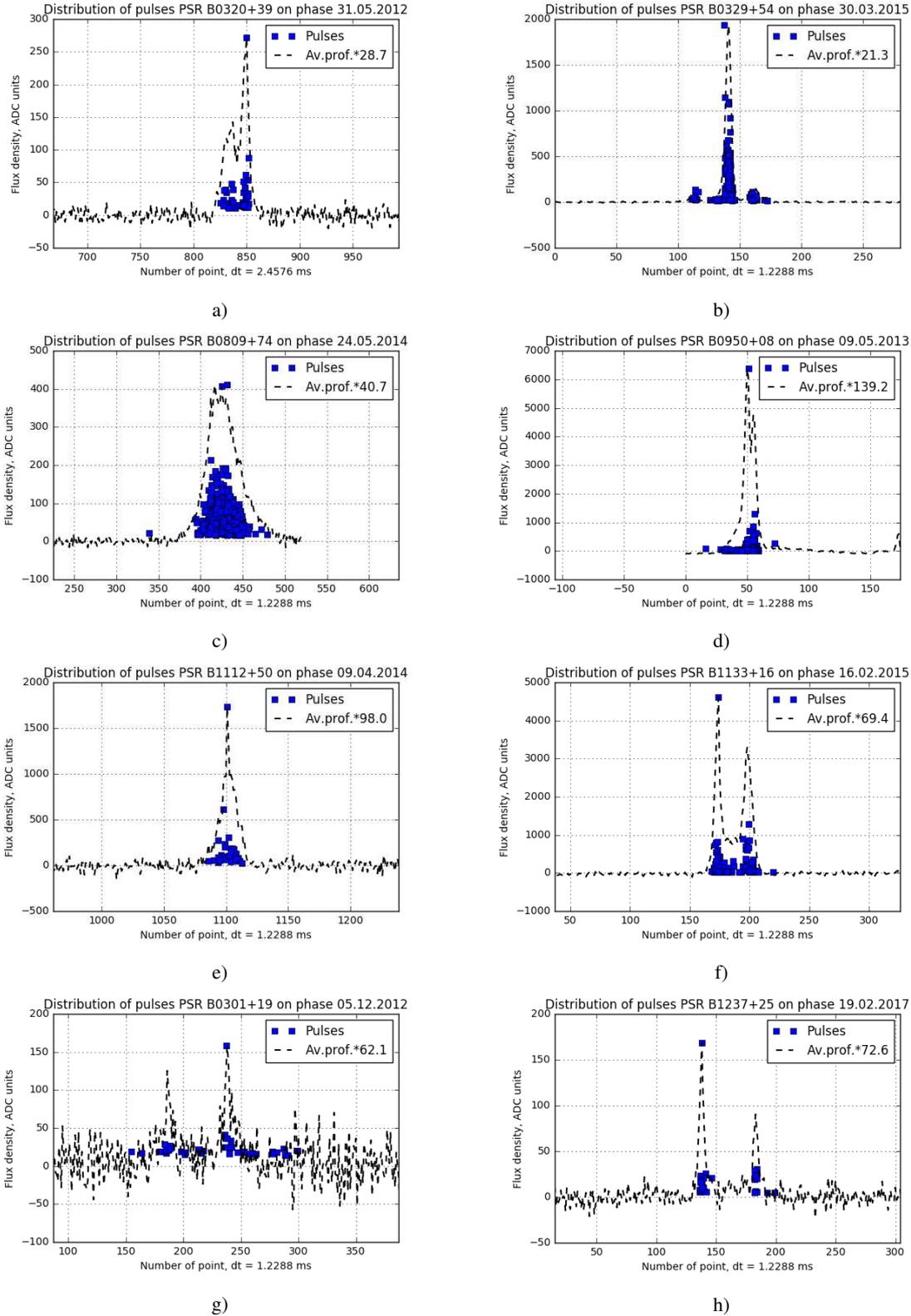}
   \caption{A distribution of individual pulses of a pulsar in a session of observation is showed with increased dynamical average profile.}
   \label{fig_disp}
\end{figure}

First report about GRPs from B1327+25 was published in (\cite{kazantsev15a}), but more detailed statistical analysis was made by \cite{kazantsev17b}. One of an example observed GRPs is shown in Figure~\ref{fig_pulse}~(h). Sample for statistical analysis was increased in about 2 times and includes 22207 pulses which are more 4$\sigma_{noise}$ by amplitude. 168 pulses more than 30$S_{avg}$ by intensity were detected. The peak flux distribution on Fig. \ref{fig_dis_energy} ~(h) demonstrates clear one-component distribution that differs from two-component distribution obtained in our earlier research both in power and form.

\begin{table}
\begin{center}
\caption[]{Peak Flux Density Distribution of GRPs and Strongest Individual Pulses Observed.}\label{Tab:publ-works}

%%Please Capitalize the First Letter of Each Notional Word in table's caption
\begin{tabular}{clccl}
  \hline\noalign{\smallskip}
No &  BName & $\alpha$ & Max pulse [$S_{avg}$]&  Max pulse [Jy] ***\\
  \hline\noalign{\smallskip}
 
1  & B0320+39 & ---     & 28.7  & 196 \\
2  & B0329+54 & ---     & 21.3  & 1680  *\\
3  & B0809+74 & ---     & 40.7  &  2680 \\
 \hline\noalign{\smallskip}
1  & B0301+19 & -2.32$\pm$0.17  & 106.5  & 687\\
2  & B0950+08 & -1.93$\pm$0.03  & 142.1  &  16800 \\
3  & B1112+50 & -2.86$\pm$0.11  & 206.5 & 1490 \\
4  & B1133+16 & -1.28$\pm$0.04/-2.78$\pm$0.06  ** & 86.7  & 5030 *   \\
5  & B1237+25 & -2.14$\pm$0.05   & 108.5  & 1350     \\
  \noalign{\smallskip}\hline
\multicolumn{5}{l}{$\alpha$ is the power of distribution of the absolute peak flux density (the slope of the straight in Log-Log scale)}\\ 
\multicolumn{5}{l}{* An upper limit due to the shortage of calibration data - see Sect. \ref{sect:analysis} for explanation. } \\
\multicolumn{5}{l}{** Estimation for 1st and 2nd components of power-law distribution.} \\
\multicolumn{5}{l}{*** Uncertainty is ~18\% for all flux estimations.}
\end{tabular}
\end{center}
\end{table}

\section{Discussion and Conclusions}
\label{sect:conclusion}

The long-time observations of 71 pulsars of Northern hemisphere were carried out at the 111 MHz with Large Phase Array in order to search of giant radio pulses. The analysis of fluctuation of pulses' intensity from B0320+39 and B0329+54 has shown an absence of pulses which can be classified as giant. Pulses' peak flux density distributions both by $S_{avg}$ and by absolute units (Jy) don't demonstrate clear bimodal form that generally referred to be typical for pulsars with GRPs. Nevertheless, the distribution for B0329+54 has complicated form that can't be described by simple log-normal distribution.

Anomalously strong pulses from B0809+74 were discovered at the 111 MHz. The pulses form a qualitatively significant power-law tail in peak flux density distribution by $S_{avg}$.

Generation of GRPs from B0301+19, B0950+08, B1112+50, B1133+16 and B1237+25 were confirmed. Statistical distribution of pulses for long-time observations were formed for these pulsars. These distributions have normally bimodal form: power-law for strong and log-normal for regular pulses.

\begin{figure}
   \centering
   \includegraphics[width=15 cm,angle=0]{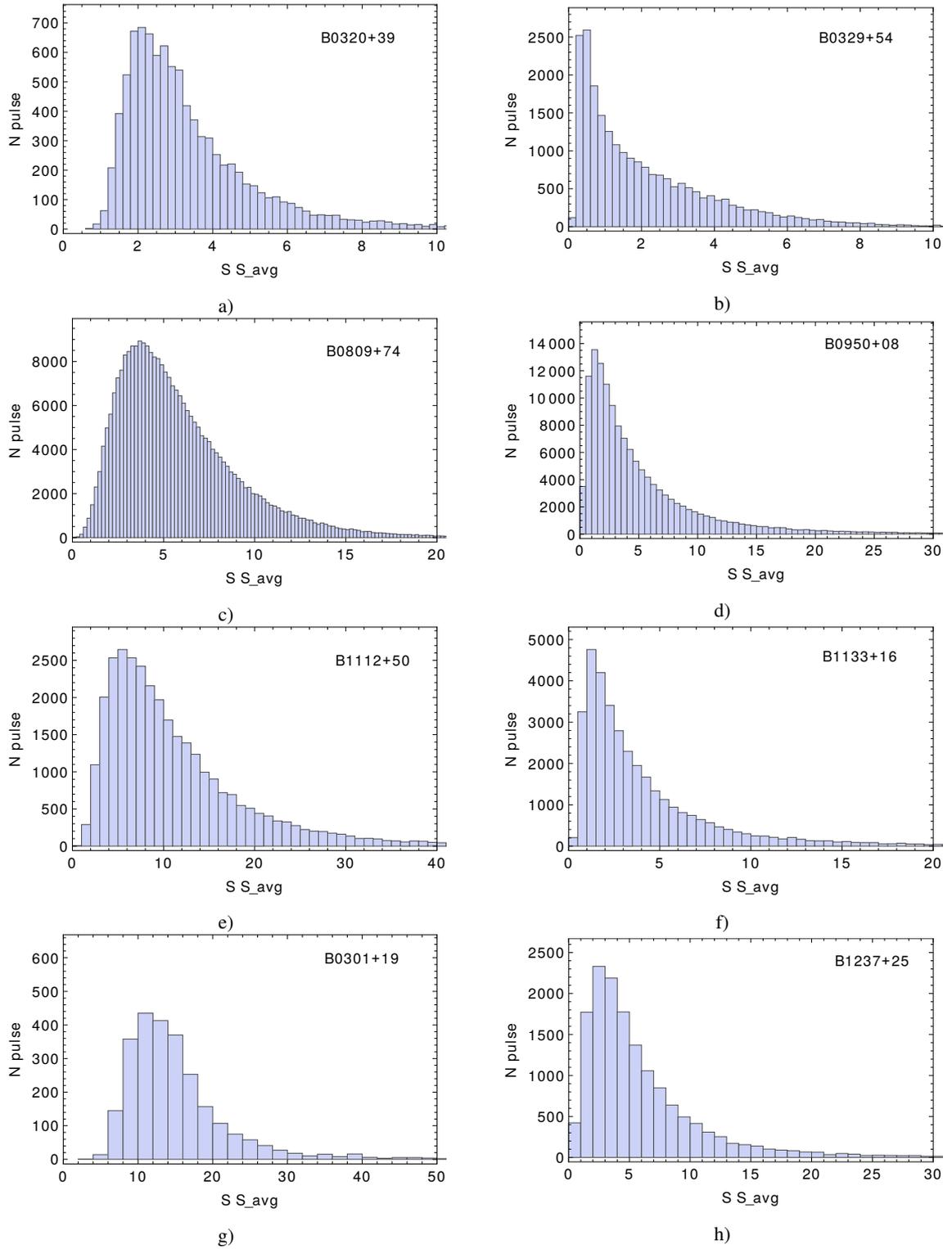}
   \caption{Pulse intensity histogram at 111 MHz for a pulsar in $S_{avg}$ units.}
   \label{fig_hist}
\end{figure}

\begin{figure}
   \centering
   \includegraphics[width=15 cm,angle=0]{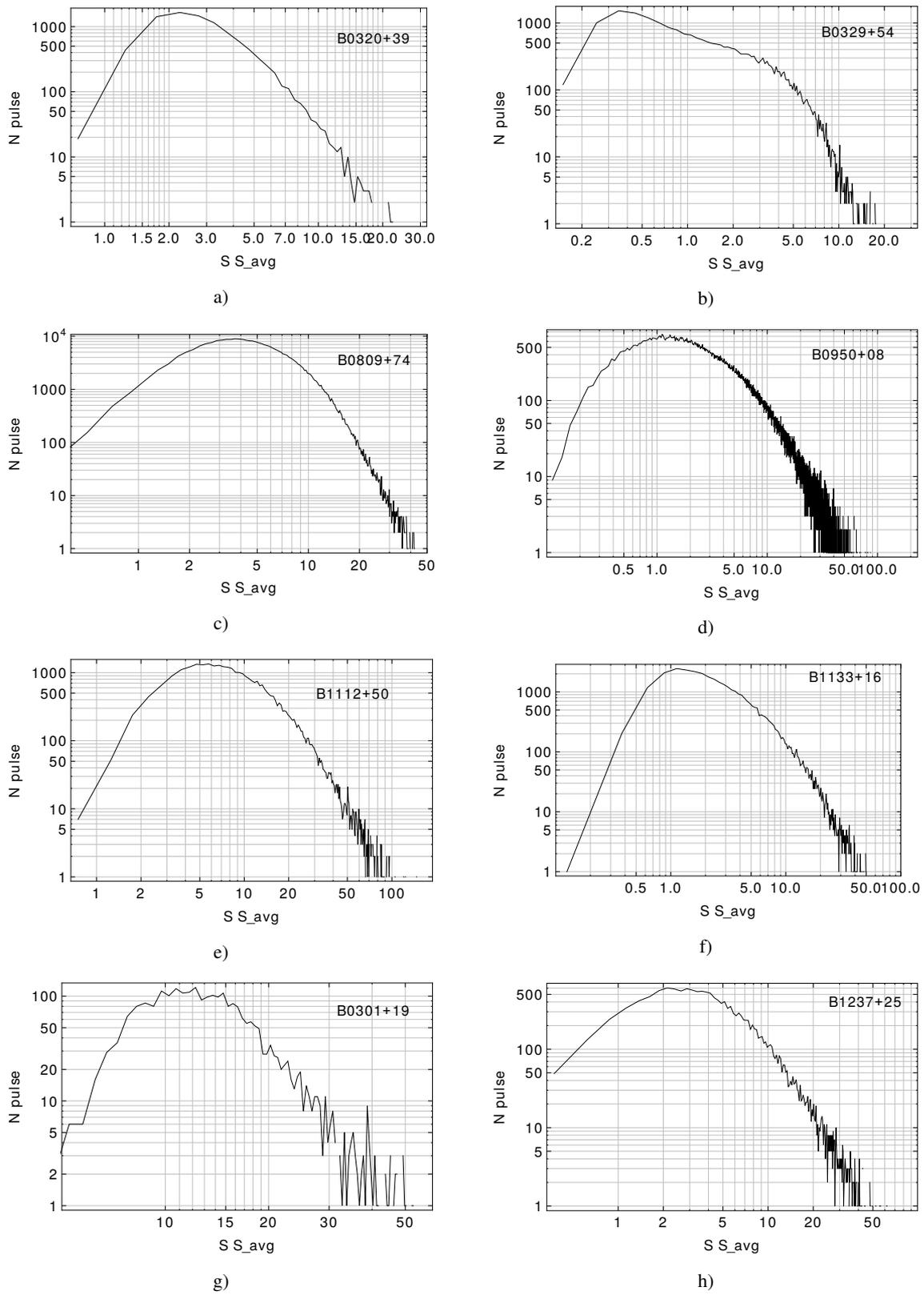}
   \caption{A distribution of the peak flux densities of individual pulses in $S_{avg}$ units.}
   \label{fig_dis_pulse}
\end{figure}

\begin{figure}
   \centering
   \includegraphics[width=15 cm,angle=0]{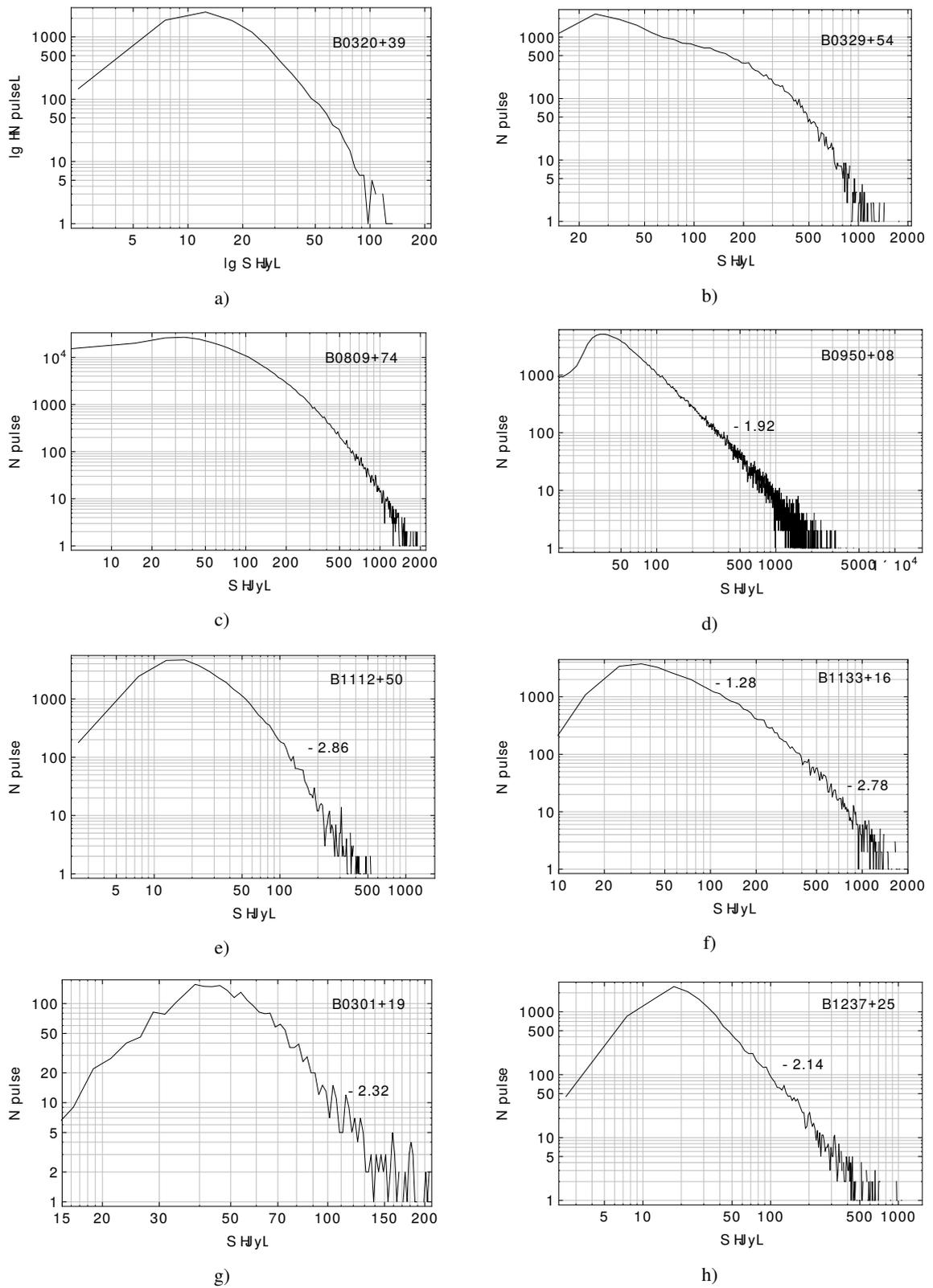}
   \caption{A distribution of the peak flux densities of individual pulses in absolute (Jy) units.}
   \label{fig_dis_energy}
\end{figure}

It is not surprising that forms of pulse intensity histograms at 111 MHz (Fig. \ref{fig_hist}) differs from the same histograms at 2695 MHz for pulsar with GRPs (see \cite{hesse74}). The fact may be explained by instability of the radio emission of the active pulsars in time. The difference between histogram at 111 MHz and at 2695 MHz for high stable pulsar without GPRs -- B0329+54 -- is much more interesting and probably may be related to the distinction of the generation of radio emission at the low and high frequencies.

It worth to note that in general we can't see quite strict qualitative differences between distributions of the "regular" pulsars and pulsars with GRPs. On the contrary, it is typical to see a "Zoo" of the distributions of quite complicated forms that gradually changes from one ("regular") class of pulses to another (GRPs). This has as the consequence of the fact that statistical criterion is not so robust in case of GRPs generated by pulsars with low magnetic field on light cylinder and should be used with caution to classify GRPs. Moreover, sporadic generation of strong pulses in dataset of such pulsars as B0329+54 and B0809+74 together with their complex and asymmetrical distribution may be caused by the fact that the some amount of pulses is generated by the same physical mechanism as GRPs, and that this phenomenon is much more typical for pulsars than it was supposed earlier.

\begin{acknowledgements}
We thank S.V. Logvinenko for technical support during the observations and V.V.Oreshko and S.A.Tyulbashev for useful discussion. This work was in part supported by the Program of the Presidium of Russian Academy of Sciences "Non-stationary processes in the Universe".
\end{acknowledgements}

\label{lastpage}

\end{document}